\documentclass[letterpaper]{jpconf}
\usepackage{graphicx}
\begin{document}
\title{Perspectives on the quantum Zeno paradox\footnote{Work of the U.S. government.  Not subject to U.S. copyright}}
\author{Wayne M. Itano}

\address{Time and Frequency Division, National Institute of Standards and Technology, Boulder, CO 80305 USA}

\ead{itano@boulder.nist.gov}

\begin{abstract}
As of October 2006, there were approximately 535 citations to the
seminal 1977 paper of Misra and Sudarshan that pointed out the
quantum Zeno paradox (more often called the quantum Zeno effect). In
simple terms, the quantum Zeno effect refers to a slowing down of
the evolution of a quantum state in the limit that the state is
observed continuously. There has been much disagreement as to how
the quantum Zeno effect should be defined and as to whether it is
really a paradox, requiring new physics, or merely a consequence of
``ordinary'' quantum mechanics.  The experiment of Itano, Heinzen,
Bollinger, and Wineland, published in 1990, has been cited around
347 times and seems to be the one most often called a demonstration
of the quantum Zeno effect. Given that there is disagreement as to
what the quantum Zeno effect {\em is}, there naturally is
disagreement as to whether that experiment demonstrated the quantum
Zeno effect. Some differing perspectives regarding the quantum Zeno
effect and what would constitute an experimental demonstration are
discussed.

\end{abstract}

\section{Introduction}

A recent entry in  Wikipedia, an Internet-based encyclopedia,
defines the quantum Zeno effect as follows:

\begin{quote}The quantum Zeno effect is a quantum mechanical phenomenon
first described by George Sudarshan and Baidyanaith Misra of the
University of Texas in 1977. It describes the situation that an
unstable particle, if observed continuously, will never decay. This
occurs because every measurement causes the wavefunction to
``collapse'' to a pure eigenstate of the measurement basis
\cite{wiki}.
\end{quote}

This definition is close to the original language of Misra and
Sudarshan \cite{misra77}, but is not sufficiently general to
describe the many situations that are considered to be examples of
the quantum Zeno effect. It is true that the quantum Zeno effect
describes the situation in which the decay of a particle can be
prevented by observations on a sufficiently short time scale.
However, the quantum Zeno effect is much more general, since it
describes the situation in which the time evolution of {\em any}
quantum system can be slowed by sufficiently frequent
``observations.''  The references to observations and to
wavefunction collapse tend to raise unnecessary questions related to
the interpretation of quantum mechanics. Actually, all that is
required is that some interaction with an external system disturb
the unitary evolution of the quantum system in a way that is
effectively like a projection operator. Finally, the word ``never''
describes a limiting case.  A slowing  of the time evolution, as
opposed to a complete freezing, is generally regarded as a
demonstration of the quantum Zeno effect.

\begin{figure}[htb]
\begin{center}
\includegraphics[width=5in]{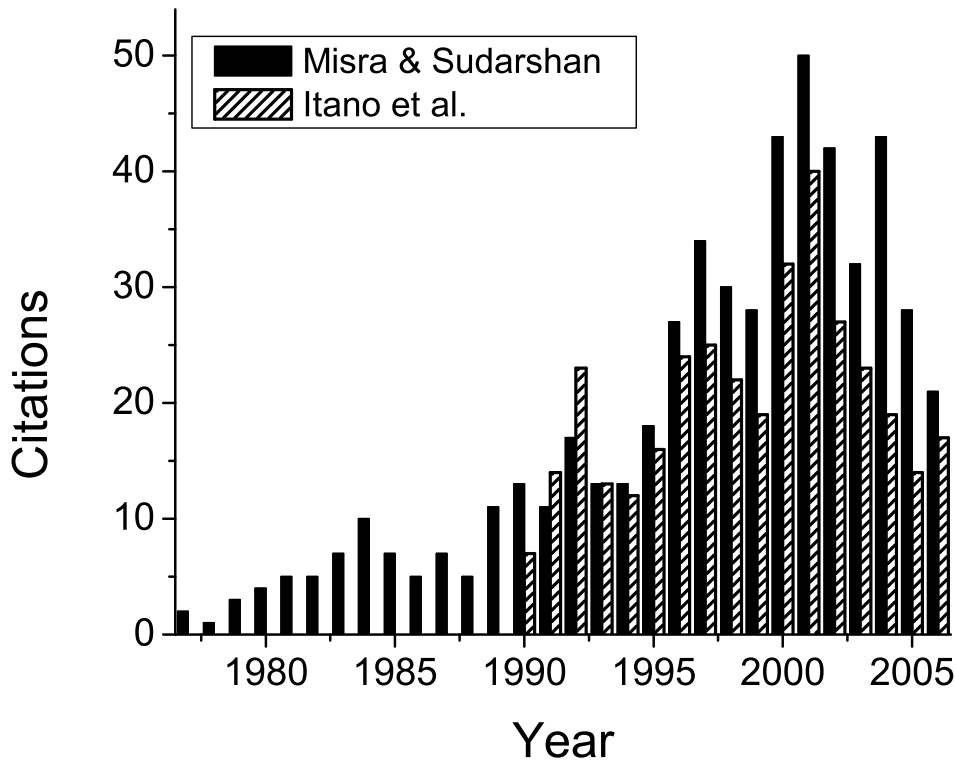}
\end{center}
\caption{\label{citesfig}Graph of the number of citations per year
to Misra and Sudarshan  \cite{misra77} and to Itano {\em et al}
\cite{itano90}.}
\end{figure}

\section{The Misra and Sudarshan Paper}

The 1977 article ``The Zeno's paradox in quantum theory'' by Misra
and Sudarshan \cite{misra77} studied the evolution of a quantum
system subjected to frequent ideal measurements. They showed that,
in the limit of infinitely frequent measurements, a quantum system
would remain in its initial state.  Applied to the case of an
unstable particle whose trajectory is observed in a bubble chamber
or film emulsion, this result seemed to imply that such a particle
would not decay, in contradiction to experiment. In this case, the
resolution to the apparent paradox lies in the fact that the
interactions between the particle and its environment that lead to
the observed track are not sufficiently frequent to modify the
particle's lifetime.

The time distribution of literature citations to Misra and Sudarshan
\cite{misra77} is shown in Fig. \ref{citesfig}.  The total number of
citations listed in the Web of Science database in October 2006 was
535.  The graph shows a relatively low but steady number of
citations per year for about a decade, followed by a large increase
that continues for over a decade, possibly peaking about 25 years
after the original publication date.  The great increase in the rate
of citations in recent years is partially due to the increased
interest in quantum information processing, where the quantum Zeno
effect may find practical applications.

\begin{figure}[hbt]
\begin{center}
\includegraphics[width=4in]{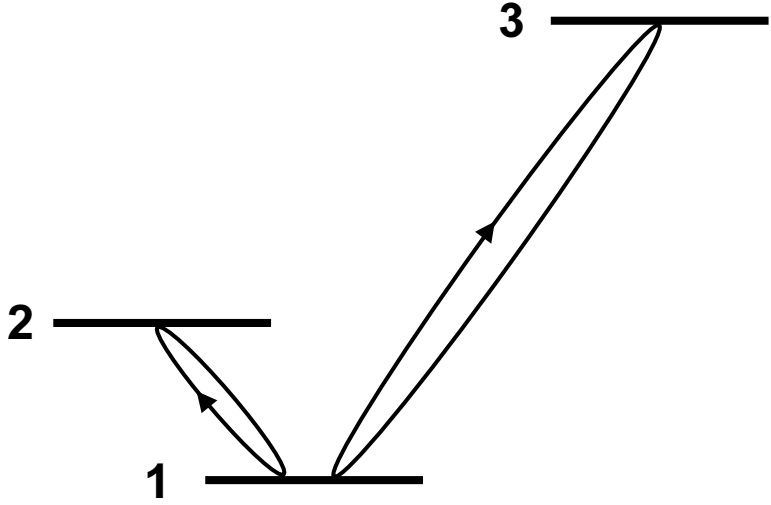}
\end{center}
\caption{\label{3levelfig}Energy-level diagram for the IHBW
demonstration of the quantum Zeno effect.}
\end{figure}

\section{Simple derivation of the quantum Zeno effect}

The quantum Zeno effect can be derived in an elementary way by
considering the short-time behavior of the state vector
\cite{peres80}.  (The treatment of Misra and Sudarshan
\cite{misra77} is more general since it involves the density
matrix.) Let $|\phi\rangle$ be the state vector at time $t=0$. If
$H$ is the Hamiltonian, in units where $\hbar=1$, then the state
vector at time $t$ is $e^{-iHt}|\phi\rangle$, and the survival
probability is $S(t)$ = $|\langle \phi|e^{-iHt}|\phi\rangle|^2$. If
$t$ is small enough, it should be possible to make a power series
expansion:
\begin{equation}
e^{-iHt} \approx  I -iHt -\frac{1}{2} H^2 t^2
\ldots,\label{quadratic}
\end{equation}
so that the survival probability is
\begin{equation}
S(t) = |\langle \phi|e^{-iHt}|\phi\rangle|^2 \approx [1-(\Delta
H)^2t^2],
\end{equation}
 where
\begin{equation}
(\Delta H)^2 \equiv \langle\phi|H^2|\phi\rangle -
\langle\phi|H|\phi\rangle^2.
\end{equation}

Many quantum systems have states whose survival probability appears
on ordinary time scales to be a decreasing exponential in time. This
is inconsistent with the quadratic time dependence of
Eq.~(\ref{quadratic}) and implies that in such cases
Eq.~(\ref{quadratic}) holds only for very short times. Consider the
survival probability $S(T)$, where the interval $[0,T]$ is
interrupted by $n$ measurements at times $T/n, 2T/n,\ldots,T$.
Ideally, these measurements are instantaneous projections and the
initial state $|\phi\rangle$ is an eigenstate of the measurement
operator. In that case, the survival probability is
\begin{equation}
S(T) \approx [1-(\Delta H)^2(T/n)^2]^n, \label{survival_n}
\end{equation}
which approaches 1 as $n\rightarrow \infty$.

It is important to note that at this level there should be nothing
controversial or problematic about the existence of the quantum Zeno
effect. The quantum Zeno effect should be observed as long as the
physical system can be made to display the behavior shown in
Eq.~(\ref{survival_n}). For a given system, it may be difficult or
impossible to make measurements quickly enough for the quadratic
time dependence of the survival probability to be observed, so that,
as a practical matter, the quantum Zeno effect cannot be observed.
It should be noted that the semantic arguments over terms such as
``measurement'' or ``observation'' can be avoided if we accept that
a ``measurement'' is an operation that interrupts the unitary time
evolution governed by $H$ in such a way as to yield
Eq.~(\ref{survival_n}) as a good approximation. That is, the
``measurement'' should effectively act as a projection operator.
According to this view, it is not necessary that the
``measurements'' be recorded by a macroscopic apparatus or that they
be instantaneous.

\begin{figure}[htb]
\begin{center}
\includegraphics[width=4in]{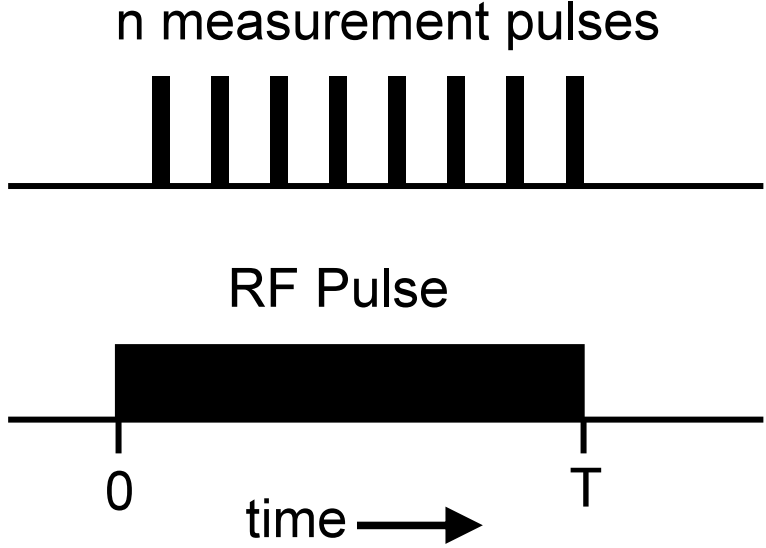}
\end{center}
\caption{\label{pulsefig}Timing of the radio frequency and optical
fields applied to the beryllium ions in the IHBW experiment}
\end{figure}

\section{The IHBW Experiment}

The experiment of Itano, Heinzen, Bollinger, and Wineland (IHBW)
\cite{itano90} was based on a proposal of Cook \cite{cook88} for
observing the quantum Zeno effect in a three-level atom (see
Fig.~\ref{3levelfig}).  Levels 1 and 2 are stable on the time scale
of the experiment.  Level 3 decays to level 1 with the emission of a
photon.  In the experiment of IHBW, levels 1 and 2 were two of the
hyperfine sublevels of the ground $^2S_{1/2}$ state of the Be$^+$
ion. Level 3 was a sublevel of the $^2P_{3/2}$ excited state that
decayed only to level 1.

The experiment was carried out with a sample of about 5000 Be$^+$
ions confined by electric and magnetic fields in a Penning trap. The
steps in the experiment were as follows:
\begin{enumerate}
\item{The ions were prepared in level 1 by optical pumping with the
laser beam.}
\item{A resonant radio frequency (RF) magnetic field was applied for
the interval required to drive the ions to level 2.}
\item{During the time that the RF pulse was applied, a variable
number $n$ of equally spaced short laser pulses was applied to the
ions (see Fig.~\ref{pulsefig}).}
\item{The laser (resonant with the 1-to-3 transition) was turned on,
and the induced fluorescence was recorded.}
\end{enumerate}

\begin{figure}[htb]
\begin{center}
\includegraphics[width=5in]{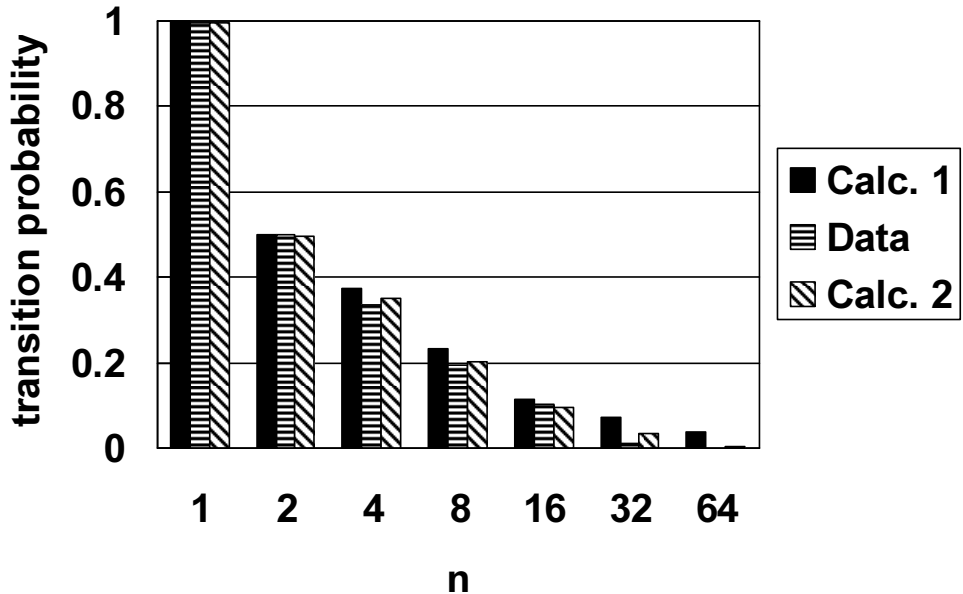}
\end{center}
\caption{\label{datafig}Probability of making the 1-to-2 transition
as a function of the number $n$ of optical ``measurement'' pulses.}
\end{figure}

The intensity of the laser-induced fluorescence at the end of the
experiment was proportional to the population of level 1. If there
are no optical pulses during the long RF pulse, the population of
level 2 as a function of the time $t$ that the RF pulse is applied
is
\begin{equation}
P_2(t)=\sin^2(\Omega t/2) = \frac{1}{2}[1-\cos(\Omega
t)],\label{rabiflop}
\end{equation}
where $\Omega$ is proportional to the amplitude of the RF field. If
the duration of the RF pulse is chosen to be $T=\pi/\Omega$ (a
pi-pulse), then all of the population is transferred from level 1 to
level 2. If $n$ equally-spaced laser pulses of negligible duration
are applied during the RF pi-pulse, the population of level 2 at
time $T$ is
\begin{equation}
P_2(T)=\frac{1}{2}\left[1-\cos^n(\pi/n)\right],\label{simplecalc}
\end{equation}
which approaches 0 as $n$ goes to infinity.

Figure \ref{datafig} compares the data to theory. The solid bars
represent the transition probability as a function of $n$ according
to the simplified calculation of Eq.~(\ref{simplecalc}). The bars
with horizontal stripes represent the data.  The bars with diagonal
stripes represent a calculation that takes into account the finite
duration of the laser pulses and optical pumping effects. The data
are in reasonably good agreement with the simplified calculation and
in better agreement with the improved calculation. The decrease in
$P_2(T)$ as $n$ increases demonstrates the quantum Zeno effect.

A variation of the experiment was carried out by initializing the
ion in level 2 and then applying the RF field and the laser pulses.
In this case, the transition from level 2 to level 1 was inhibited
as $n$ increased.  This is another example of the quantum Zeno
effect.  In this case, the inhibition of the transition is
accompanied by the {\em absence} of laser-induced fluorescence.

Recently, the quantum Zeno effect was observed for an unstable
quantum system by Fischer {\em et al} \cite{fischer01}. The quantum
Zeno effect for induced transitions and for unstable systems are not
fundamentally different, since they both follow from the general
arguments of Misra and Sudarshan \cite{misra77}, but it has been
difficult to observe in the latter case, because of the short times
over which the decay is nonexponential. Fischer {\em et al} were
able to create an artificial system (atoms tunneling from a
standing-wave light field) in which the interactions could be
controlled so as to observe the desired effects.

\section{Responses to the IHBW Experiment}

As can be seen by the history of citations (Fig.~\ref{citesfig}),
the publication of the IHBW experiment \cite{itano90} generated
considerable interest.  Initially, some of the responses were
critical in one way or another.  Some (e. g.,
Ref.~\cite{ballentine91}) objected to the use of the term
``wavefunction collapse'' in describing the experiment.  The authors
responded that the concept of wavefunction collapse was not
essential, and that any interpretation of quantum mechanics that
yielded the same prediction of the experimental results should be
regarded as valid \cite{itano91}.  Some objected to the fact that
photons were not actually observed during the intermediate
``measurements,'' in the sense of having the scattered photons
registered by a detector, so that the experiment did not actually
demonstrate the quantum Zeno effect
\cite{ballentine91,petrosky90,inagaki92,pascazio93}. However, the
results are predicted to be the same whether or not the intermediate
measurements are made. It is enough that the measurements {\em
could} have been made.   As long as the laser interactions act
effectively as projection operators, so that the algebra of
Eqs.~(\ref{quadratic})--(\ref{survival_n}) is followed, the
experiment should be regarded as a demonstration of the quantum Zeno
effect. It should be noted that none of the criticisms were directed
at the execution of the experiment itself, only at the
interpretation. For the most part, the citations to
Ref.~\cite{itano90} simply accept it as a demonstration of the
quantum Zeno effect.  In fact, it is cited in quantum mechanics
textbooks \cite{griffiths95,sakurai94,peres93,braginsky92} and
popular science books \cite{whitaker96,coveney90,schulman97}.

\section{Distinctions}

While Misra and Sudarshan originally used the term ``quantum Zeno
paradox,'' as did Peres \cite{peres80} and others, the more recent
work usually uses the term ``quantum Zeno effect,'' perhaps because
the effect no longer seems paradoxical.  Some authors distinguish
between the quantum Zeno {\em paradox} and the quantum Zeno {\em
effect}, but they do so in differing ways. Pascazio and Namiki
\cite{pascazio94} call the situation in which the frequency of
measurements is finite and the evolution is slowed the quantum Zeno
{\em effect,} and the limiting case in which the frequency of
measurements is infinite and the evolution is frozen the quantum
Zeno {\em paradox}.   Block and Berman \cite{block91} call the
inhibition of spontaneous decay the quantum Zeno {\em paradox} and
the inhibition of induced transitions (as in the IHBW experiment)
the quantum Zeno {\em effect}. In Ref. \cite{home92}, Home and
Whitaker reserve the term quantum Zeno {\em paradox} for a
negative-result experiment involving observations with a macroscopic
apparatus.  This definition of the quantum Zeno paradox seems to
exclude most, if not all, feasible experiments.  In this context,
the IHBW experiment is not regarded as an example of the quantum
Zeno paradox because a local interaction is present between the
laser field and the atoms, and also because the electromagnetic
field, containing zero or a few scattered photons, is not regarded
as a macroscopic observation apparatus.  They regard the type of
experiments where the time evolution of a quantum system is affected
by a direct interaction, for example with an external field, as
examples of the quantum Zeno {\em effect}.  However, in a later
publication \cite{home97} the same authors treat the terms quantum
Zeno {\em paradox} and quantum Zeno {\em effect} as synonymous and
restrict both to nonlocal negative-result experiments involving a
macroscopic observation apparatus.  Experiments that do not meet
these criteria would not be examples of {\em either} the quantum
Zeno paradox {\em or} the quantum Zeno effect, according to their
later definition.

\section{Extensions}

Several variations on the general theme of quantum Zeno effects have
been described.   Soon after the IHBW experiment was carried out,
Peres and Ron \cite{peres90} showed that a partial quantum Zeno
effect results if the measurements are too weak to completely
destroy the coherence of the state of the measured system.  A
modification of the IHBW experiment was proposed in which the
measurement laser pulses are weakened.  Jordan {\em et al}
\cite{jordan91} showed that a related effect, damped oscillations of
the state populations, can occur if the duration of the experiment
is extended, while weak measurements are made.

Some, including Kofman and Kurizki \cite{kofman00} and Facchi {\em
et al} \cite{facchi01} have shown that the decay of an unstable
quantum system can be accelerated by frequent observations.  This is
called the quantum anti-Zeno effect or the inverse quantum Zeno
effect.  As is the case for the quantum Zeno effect, the
observations must take place before the decay becomes exponential.
Unlike the quantum Zeno effect, which follows from rather general
arguments, e. g. Eqs.~(\ref{quadratic})--(\ref{survival_n}), the
possibility of observing a quantum anti-Zeno effect depends on the
details of the system. The experiment of Fischer {\em et al}
\cite{fischer01} demonstrated the quantum anti-Zeno effect as well
as the quantum Zeno effect. 

An interesting generalization of the quantum Zeno effect is the
concept of quantum Zeno dynamics \cite{facchi00,facchi02}.  Frequent
measurements can confine the evolution of a quantum system to a
subspace of the Hilbert space rather than simply to the initial
state.  Compared to the ordinary quantum Zeno effect, the difference
is that the measurements distinguish not between the initial {\em
state} and all other states but between a {\em subspace} and the
rest of the Hilbert space. This form of quantum Zeno effect may find
application in quantum information processing.

\section{Applications}

As already noted, the recent increase in the rate of citations to
the articles of Misra and Sudarshan \cite{misra77} and IHBW
\cite{itano90} is partially related to increased interest in quantum
information processing.  In this context, there have been various
proposals to use the quantum Zeno effect to preserve quantum systems
from decoherence.

Beige {\em et al} \cite{beige00} have proposed an arrangement of
atoms inside an optical cavity capable of carrying out quantum logic
operations with low error rates within a decoherence-free subspace
of the Hilbert space. States outside the decoherence-free subspace
are coupled strongly to the environment. The quantum Zeno effect
then leads to effective confinement of the system to the
decoherence-free subspace, which is an example of a quantum Zeno
subspace.

Franson {\em et al} \cite{franson04} have proposed use of the
quantum Zeno effect to suppress errors in a linear optics
implementation of quantum computation. In this implementation, the
presence of two photons in the same mode indicates an error. The
presence of a strong two-photon absorber in an optical fiber takes
the role of the ``observer'' and suppresses the errors. Other
proposed applications of the quantum Zeno effect to error prevention
in quantum computation are discussed in Refs.
\cite{zurek84,vaidman96,pachos02,ralph03}.

Quantum ``bang-bang'' control and related dynamical decoupling
techniques \cite{viola98,viola99} utilize frequent, pulsed
interactions to effectively prevent decoherence of a quantum system
by confining the dynamics to a subspace.  This is not exactly the
quantum Zeno effect, since the interactions are unitary, but the
results are mathematically similar to those for the quantum Zeno
effect.

Dhar {\em et al} \cite{dhar06} have discussed the ``super-Zeno
effect,'' which preserves a state (or more generally, keeps a
quantum system within a subspace of the Hilbert space) with a set of
pulsed interactions unequally spaced in time.  The timing of these
interactions can be arranged so as to be more efficient than can be
done with the same number of equally spaced interactions (ordinary
quantum Zeno effect).  Also, it should be noted that the pulsed
interactions are unitary kicks, as in the so-called ``bang-bang
control'' \cite{viola98}, and not observations in the usual sense.

\section{Conclusion}

The 1977 publication of Misra and Sudarshan stimulated a great deal
of theoretical and experimental work that has enhanced our
understanding of the time development of quantum systems, such as
the short-time nonexponential decay of unstable quantum systems. The
results of the IHBW experiment, published in 1990, was a clear
confirmation of the existence of the quantum Zeno effect for the
case of the inhibition of an induced transition.  Interest in the
quantum Zeno effect continues to be high, partially due to the
possibility of practical applications in quantum information
processing.


\section*{References}

\begin{thebibliography}{99}

\bibitem{wiki}\begin{verbatim}
URL:http://en.wikipedia.org/wiki/Quantum_Zeno_effect\end{verbatim}

\bibitem{misra77}Misra B and Sudarshan E C G 1977 {\it J. Math.
Phys.} {\bf 18} 756

\bibitem{itano90}Itano W M, Heinzen D J, Bollinger J J, and Wineland
D J 1990 {\it Phys. Rev.} A {\bf 41} 2295

\bibitem{peres80}Peres A 1980 {\it Am. J. Phys.} {\bf 48} 931

\bibitem{cook88}Cook R J 1988 {\it Physica Scripta} {\bf T21} 49

\bibitem{fischer01}Fischer M C, Guti\'errez-Medina B, and Raizen M G
2001 {\it Phys. Rev. Lett.} {\bf 87} 040402

\bibitem{ballentine91}Ballentine L E 1991 {\it Phys. Rev.} A {\bf
43} 5165

\bibitem{itano91}Itano W M, Heinzen D J, Bollinger J J, and Wineland
D J 1990 {\it Phys. Rev.} A {\bf 43} 5168

\bibitem{petrosky90}Petrosky T, Tasaki S, Prigogine I 1990 {\it
Phys. Lett.} A {\bf 151} 109

\bibitem{inagaki92}Inagaki S, Namiki M, and Tajiri T 1992 {\it Phys.
Lett.} A {\bf 166} 5

\bibitem{pascazio93}Pascazio S, Namiki M, Badurek G, and Rauch H
1993 {\it Phys. Lett.} A {\bf 179} 155

\bibitem{griffiths95}Griffiths D J 1995 {\it Introduction to Quantum
Mechanics} (Englewood Cliffs, NJ: Prentice-Hall) p 385

\bibitem{sakurai94}Sakurai J J 1994 {\it Modern Quantum Mechanics}
Rev. Ed., ed S F Tuan (Reading, MA: Addison-Wesley) p 485

\bibitem{peres93}Peres A 1993 {\it Quantum Theory: Concepts and Methods}
(Dordrecht: Kluwer Academic) p 394

\bibitem{braginsky92}Braginsky V B and Khalili F Ya 1992 {\it
Quantum Measurement} ed K S Thorne (Cambridge: Cambridge) p 94

\bibitem{whitaker96}Whitaker A 1996 {\it Einstein, Bohr and the
Quantum Dilemma} (Cambridge: Cambridge) p 313

\bibitem{coveney90}Coveney P and Highfield R 1990 {\it The Arrow of
Time} (New York: Fawcett Columbine) p 137

\bibitem{schulman97}Schulman L S 1997 {\it Time's Arrows and Quantum
Measurement} (Cambridge: Cambridge) p 64

\bibitem{pascazio94}Pascazio S and Namiki M 1994 {\it Phys. Rev.} A
{\bf 50} 4582

\bibitem{block91}Block E and Berman P R 1991 {\it Phys. Rev.} A {\bf
44} 1466

\bibitem{home92}Home D and Whitaker M A B 1992 {\it J. Phys.} A {\bf
25} 657

\bibitem{home97}Home D and Whitaker M A B 1997 {\it Annals of
Physics} {\bf 258} 237

\bibitem{peres90}Peres A and Ron A 1990 {\it Phys. Rev.} A {\bf 42}
5720

\bibitem{jordan91}Jordan T F, Sudarshan E C G, and Valanju P 1991 {\it
Phys. Rev.} A {\bf 44} 3340

\bibitem{kofman00}Kofman A G and Kurizki G 2000 {\it Nature} {\bf
405} 546

\bibitem{facchi01}Facchi P, Nakazato H, and Pascazio S 2001 {\it Phys.
Rev. Lett.} {\bf 86} 2699

\bibitem{facchi00}Facchi P, Gorini V, Marmo G, Pascazio S, and
Sudarshan E C G 2000 {\it Phys. Lett.} A {\bf 275} 12

\bibitem{facchi02}Facchi P and Pascazio S 2002 {\it Phys. Rev.
Lett.} {\bf 89} 080401

\bibitem{beige00}Beige A, Braun D, Tregenna B, and Knight P L 2000
{\it Phys. Rev. Lett.} {\bf 85} 1762

\bibitem{franson04}Franson J D, Jacobs B C, and Pittman T B 2004
{\it Phys. Rev.} A {\bf 70} 062302

\bibitem{zurek84}Zurek W H 1984 {\it Phys. Rev. Lett.} {\bf 53} 391

\bibitem{vaidman96}Vaidman L, Goldenberg L, and Wiesner S {\it Phys.
Rev.} A {\bf 54} R1745

\bibitem{pachos02}Pachos J and Walther H 2002 {\it Phys. Rev. Lett.}
{\bf 89} 187903

\bibitem{ralph03}Ralph T C, Gilchrist A, Milburn G J, Munro W J, and
Glancy S 2003 {\it Phys. Rev.} A {\bf 68} 042319

\bibitem{viola98}Viola L and Lloyd S 1998 {\it Phys. Rev.} A
{\bf 58} 2733

\bibitem{viola99}Viola L, Knill E, and Lloyd S 1999 {\it Phys. Rev.
Lett.} {\bf 82} 2417

\bibitem{dhar06}Dhar D, Grover L K, and Roy S M 2006 {\it Phys. Rev.
Lett.} {\bf 96} 100405



\end{thebibliography}
\end{document}